\documentclass[aps,prl,twocolumn,groupedaddress]{revtex4-1}
\usepackage{graphicx}

\begin{document}


\title{Comparison of hot-electron transmission in ferromagnetic Ni on epitaxial and polycrystalline Schottky interfaces} 

\author{S. Parui,$^1$ K. G. Rana,$^1$ L. Bignardi,$^2$ P. Rudolf,$^2$ B. J. van Wees,$^1$ and T. Banerjee$^1$}
\affiliation{$^1$Physics of Nanodevices, Zernike Institute for Advanced Materials, University of Groningen, Nijenborgh 4, 9747 AG Groningen, The Netherlands}
\affiliation{$^2$Surfaces and Thin Films, Zernike Institute for Advanced Materials, University of Groningen, Nijenborgh 4, 9747 AG Groningen, The Netherlands}
\date{\today}

\begin{abstract}
The hot-electron attenuation length in Ni is measured as a function of energy across two different Schottky interfaces viz. a polycrystalline Si(111)/Au and an epitaxial Si(111)/NiSi$_2$ interface using ballistic electron emission microscopy (BEEM). For similarly prepared Si(111) substrates and identical Ni thickness, the BEEM transmission is found to be lower for the polycrystalline interface than for the epitaxial interface. However, in both cases, the hot-electron attenuation length in Ni is found to be the same. This is elucidated by the temperature-independent inelastic scattering, transmission probabilities across the Schottky interface, and scattering at dissimilar interfaces.

\vspace{1pc}
\vspace{1pc}
\end{abstract}

\maketitle
\section{\label{sec:level1}I. Introduction}
The fundamental scattering mechanisms governing hot-electron spin transport in magnetic materials over a wide energy regime is a subject of intense research. \cite{Rippard,Vlutters,Parkin,TamalikaFe,TamalikaPRL,Jansen,Heindl,Stollenwerk,Lu} Experimental techniques or devices that employ hot-electron transport, generally comprise of a  Schottky barrier at a metal-semiconductor (M/S) interface and a ferromagnetic (FM) metal base into which hot electrons are injected. Hot-electron transmission in the FM layer depends exponentially on its thickness and is influenced by the choice of the M/S interface, which acts both as an energy and momentum filter. The scattering processes during transmission depend on the energy-dependent inelastic mean-free path, elastic and quasielastic scattering processes, as well as on the group velocity of the states above the Fermi level into which the hot electrons propagate. \cite{Vlutters,TamalikaPRL}\\
\indent From transport experiments with hot electrons, the energy dependent attenuation length in several FMs such as Co, NiFe, CoFe etc. has been determined. \cite{Parkin, Vlutters, TamalikaFe, Jansen, Rippard, TamalikaPRL} The sensitivity of the hot-electron attenuation length to the momentum distribution of the injected electrons in Co was also demonstrated. \cite{Jansen} Using identically prepared metal base, it was seen that the attenuation length in Co was found to be a factor of 2 larger for hot electron injection across an amorphous Al$_2$O$_3$ tunnel barrier as compared to an ideal vacuum tunnel barrier.\\ 
\begin{figure}[b]
\includegraphics[scale=0.46]{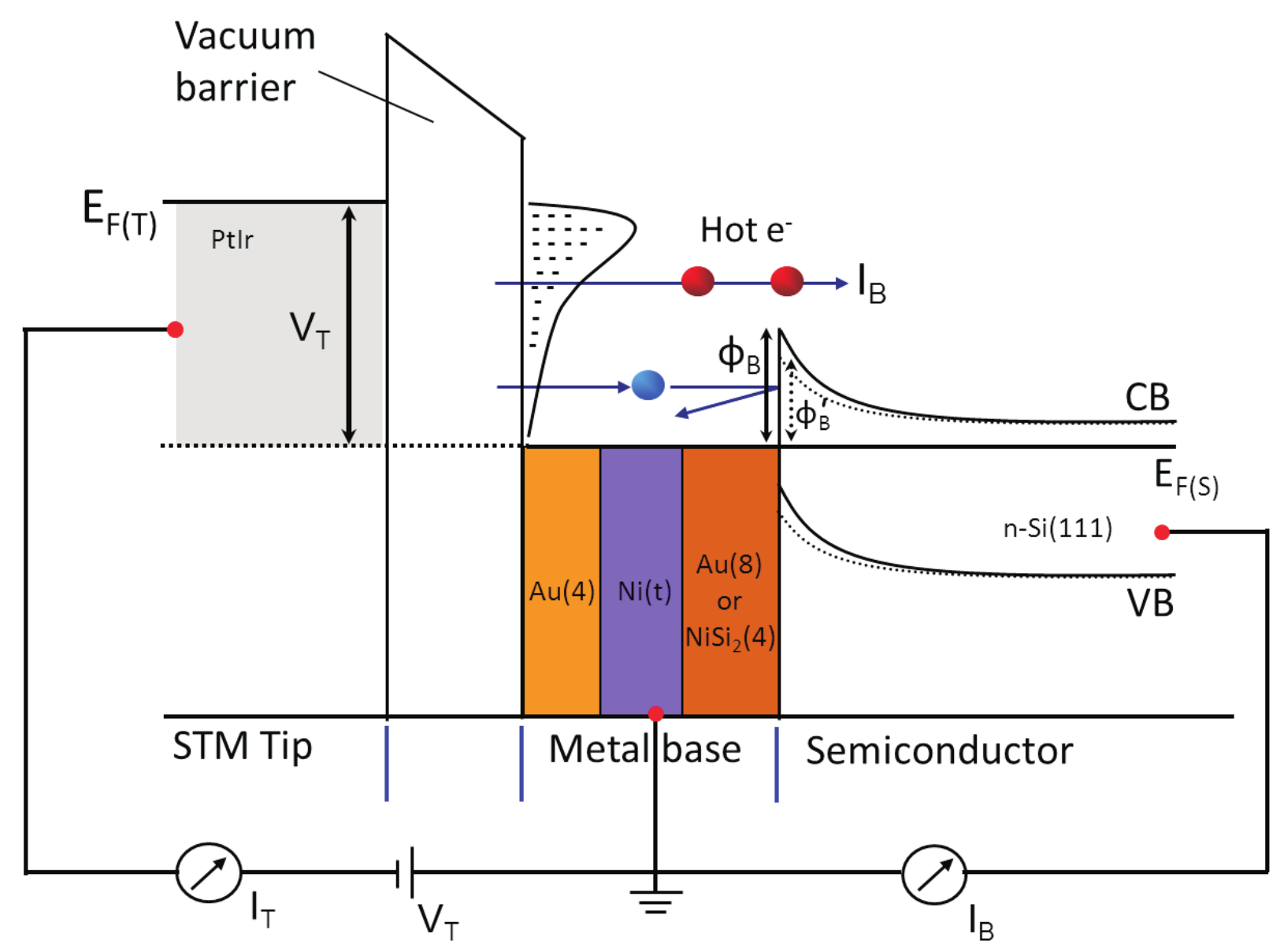}
\caption{\label{fig:Schematic} Schematic layout and energy band diagram of a BEEM experiment. A PtIr tip is used to inject hot electrons over the vacuum barrier, in a Au(NiSi$_2$)/Ni/Au base. The transmitted electrons are then collected in an n-Si(111) semiconductor. Au and NiSi$_2$ layers on n-Si(111) give rise to two different SBHs $\phi_B$ or $\phi'_B$ as depicted.
}
\end{figure}
\indent In spite of the aforementioned studies, our overall understanding of the processes governing hot-electron transmission and attenuation length in FM metals is still incomplete. In this work we investigate the energy dependence of the hot-electron attenuation length in Ni and extricate the influence of different scattering processes on it by using two different Schottky interfaces, namely, a polycrystalline Si(111)/Au and an epitaxial Si(111)/NiSi$_2$ Schottky interface. We employ Ballistic Electron Emission Microscopy (BEEM) for this purpose. Identical Si(111) surface preparation and deposition conditions for Ni ensured a similar injection profile of the hot electrons injected into the metal layers. We find that the BEEM transmission is lower in Ni across the polycrystalline Si/Au interface for all Ni thicknesses studied. However, the hot-electron attenuation length in Ni is measured to be the same for both the interfaces. Our work highlights the dominant contribution of inelastic volume scattering to the hot-electron transmission in such ferromagnetic thin films.\\
\indent Based on hot electrons, Ballistic Electron Emission Microscopy (BEEM) is an ideal technique to directly measure the attenuation lengths in thin metal films. BEEM, developed by Bell and Kaiser in 1988 \cite{KaiserBell}, is the three terminal extension of a Scanning Tunneling Microscopy (STM) where a tip (emitter) is used to inject non-equilibrium electrons (energy few eV above the Fermi level) across a vacuum-tunneling barrier into a metallic overlayer (base) deposited on top of a Schottky interface (collector). Depending on the scattering in the base, a fraction of the injected hot electrons propagates to the M/S interface and are collected as BEEM current $I_B$ when the necessary energy and momentum criteria to overcome the Schottky barrier height (SBH) are satisfied (Fig. 1). From $I_B$, measured in different regions, the local SBH can be determined using the Bell-Kaiser (BK) model. \cite{KaiserBell} By fitting the decrease in $I_B$ with metal layer thickness, it is possible to extract the hot electron attenuation length in different materials. \cite{Heindl, Jansen, Stollenwerk, Rippard, TamalikaPRL, Parui, Bell}\\
\indent In this work we investigated the hot-electron attenuation length in ferromagnetic Ni films using two different sandwich structures of Au/Ni(t)/Au and NiSi$_2$/Ni(t)/Au with Ni thickness varying from 2 - 10 nm. The top Au layer provides a chemically inert surface for \textit{ex situ} sample transfer, and the bottom Au and NiSi$_2$ layers on Si(111) form the polycrystalline and epitaxial Schottky interfaces, respectively. The choice of the two Schottky interfaces was determined by the fact that Au on Si is the canonical polycrystalline Schottky interface for hot electron studies, whereas the epitaxial Schottky interface of NiSi$_2$ on Si acts as a good energy and momentum filter for hot electrons, as established by our earlier work. \cite{Subir} $I_B$ was recorded at a fixed injection current, over a wide energy range and for each Ni thickness, on both the Schottky interfaces. The exponential decay of $I_B$ with varying Ni thickness allowed us to extract the hot-electron attenuation length in Ni at different energies.\\
\section{\label{sec:level2}II. Experiment}  
\indent Our device structures involved the preparation of atomically flat n-Si(111) substrates onto which the metal layers were deposited using an ultrahigh-vacuum molecular beam epitaxy (UHV MBE) system. Substrates consisted of buffered hydrofluoric acid (HF)-etched n-Si(111) with a lithographically defined area of 150 $\mu$m diameter surrounded by a thick oxide insulator. The Si surface was H-terminated using 1\% HF followed by a second anisotropic etching step using 40$\%$ ammonium fluoride (NH$_4$F) solution. This anisotropic etching step resulted in well-terminated Si(111) planes at the surface.\cite{Dumas} The substrates were then immediately loaded into the MBE system at a base pressure of 10$^{-10}$ mbar. Two different sets of samples were fabricated: for the first set, a thin epitaxial NiSi$_2$ was grown on an atomically flat Si(111) substrate following the well-established method described elsewhere. \cite{Tung, Subir} Further deposition of the Ni thin films of varying thicknesses (t) and of the Au cap layer (4 nm) were performed at RT. For the second set, a 8 nm Au layer was grown to form the polycrystalline Schottky interface on Si(111). Here, too, Ni layers of varying thicknesses (t) were deposited, followed by a 4 nm Au cap at RT. The thickness of each deposited layer was determined with a quartz crystal monitor and also characterized separately using glancing-incident X-ray analysis and Atomic Force Microscopy (AFM). The devices were then transferred into an UHV STM system for transport measurements.
\begin{figure}[t]
\includegraphics[scale=0.46]{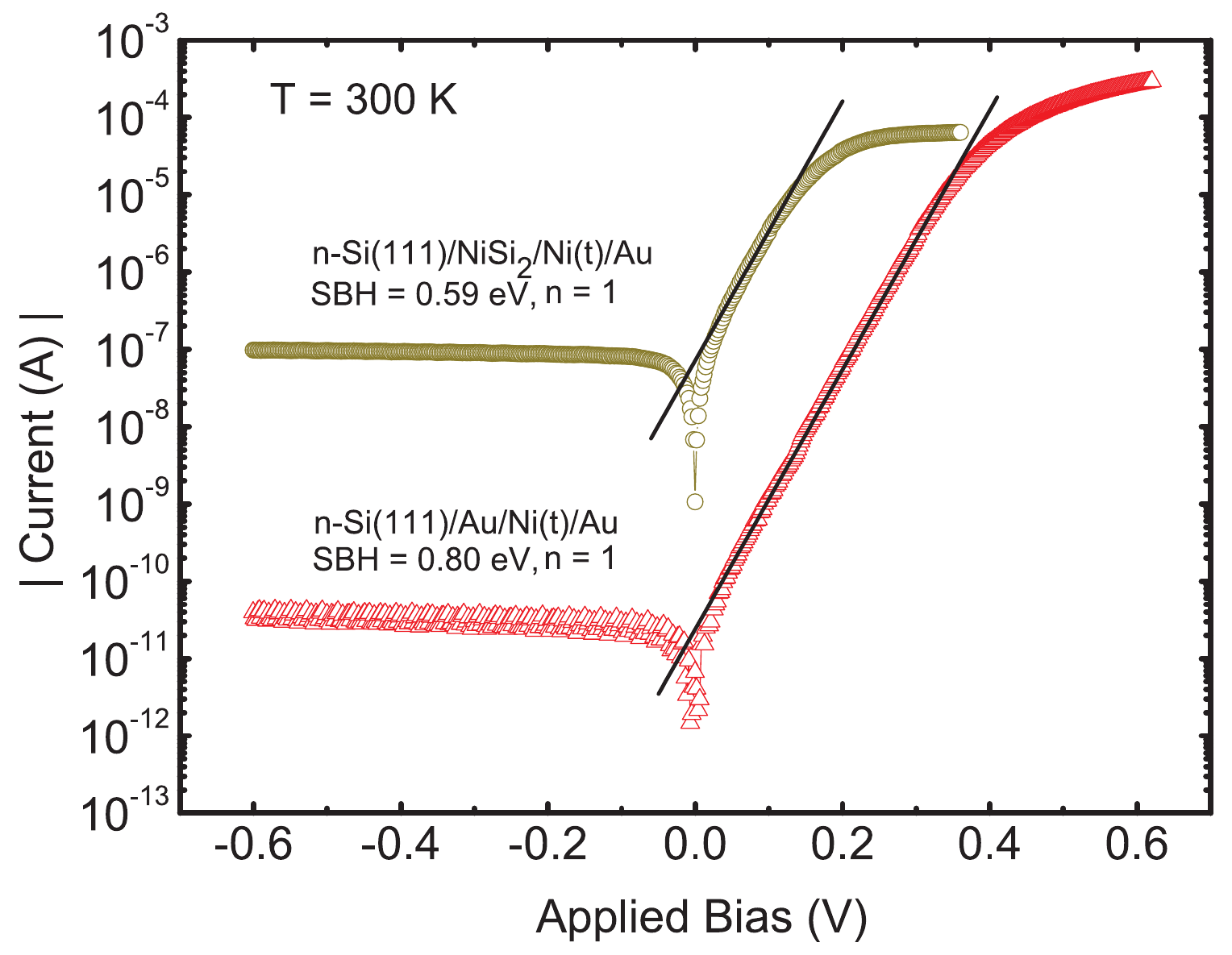}
\caption{\label{fig:Two probe} Electrical (I-V) characterization of the devices on n-Si(111) with two different Schottky barrier heights at room temperature. The active device area was 150 $\mu$m in diameter. The straight lines were fitted using the thermionic emission theory as described in the text.}
\end{figure}
\section{\label{sec:level3}III. Results}
\indent The rectifying behavior of the diodes was first characterized using standard current-voltage (I-V) measurements. Figure 2 shows a typical I-V characteristic for both the diodes. The Schottky barrier heights were obtained from the I-V plot by fitting the forward bias characteristics using thermionic emission theory:
\begin{equation}
I=A^{**}AT^2exp\left(-\frac{q\phi_B}{k_BT}\right)\left[exp\left(\frac{qV}{nk_BT}\right)-1\right]
\end{equation}
The symbols have their usual meanings \cite{Sze}. $\phi_B$ was determined to be 0.80 $\pm$ 0.02 eV and 0.59 $\pm$ 0.02 eV for the n-Si(111)/Au/Ni(t)/Au and the n-Si(111)/NiSi$_2$/Ni(t)/Au device structures, respectively with an ideality factor of n=1 for both cases.\\
\indent For the BEEM studies, a modified commercial STM system from RHK technology was used. The top metal layer of the device structure was grounded and a large area ohmic contact to the back of the n-Si(111) substrate was used to collect the transmitted electrons, $I_B$. BEEM measurements were performed at RT for all device structures with the Si(111)/Au Schottky interface and at 100 K for diodes with the Si(111)/NiSi$_2$ Schottky interface. The latter was necessary because for diodes with low SBH, as in Si(111)/NiSi$_2$ interface, Johnson noise can dominate the collected current.\\
\indent STM topography and simultaneously acquired BEEM images were obtained with a mechanically cut Pt$_{0.8}$Ir$_{0.2}$ tip at a bias of -1.4 V and  tunneling current of 1 nA as shown in Figs. 3(a), 3(d) and Figs. 3(b), 3(e), respectively for two different devices on both the M/S interfaces. The STM topography shows the morphology of the Au grains of the top metal layer for both device structure. The root mean square (rms) roughness on the surface of Au grains is found to be $\sim$0.9 nm for both the devices. The device structure on the epitaxial M/S interface has smaller Au grains ($\sim$10 nm diameter) compared to that on the polycrystalline M/S interface ($\sim$20 nm diameter). Corresponding BEEM images in Figs. 3(b) and 3 (e) represent spatial maps of the transmitted current. A quantitative analysis of the histogram of the transmitted current for both cases at $V_T$=-1.4 V is shown in Figs. 3(c) and 3(f). The mean value of $I_B$ is $2.1\pm0.5$ pA/nA for n-Si(111)/NiSi$_2$/Ni(8 nm)/Au and $0.5\pm0.2$ pA/nA for n-Si(111)/Au/Ni(8 nm)/Au.\\
\begin{figure}
\includegraphics[scale=0.49]{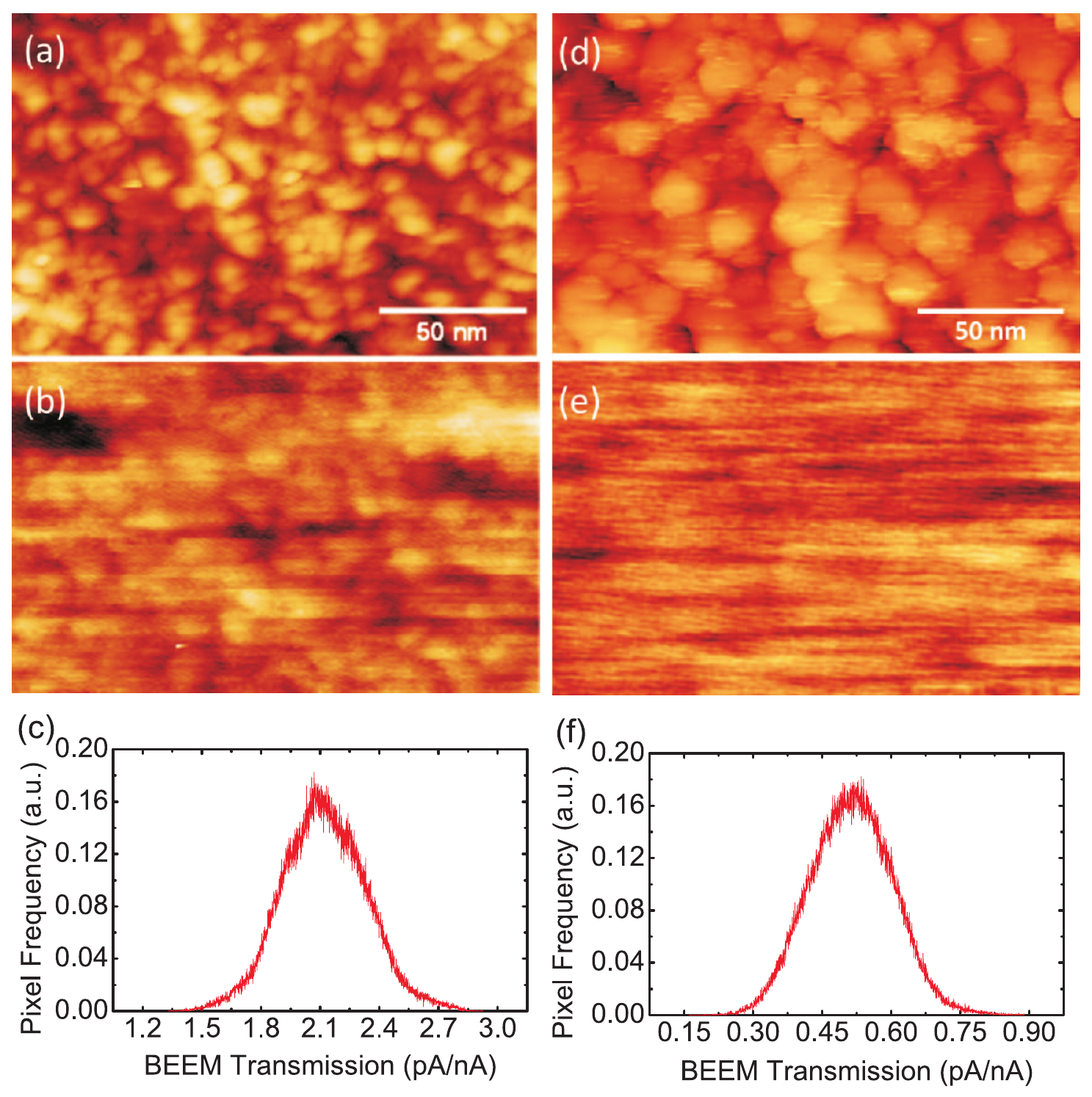}
\caption{\label{fig:BEEM} Surface topography as obtained by STM (top panels) and simultaneously recorded BEEM image (middle panels) of n-Si(111)/NiSi$_2$/Ni(8 nm)/Au [(a), (b)] and n-Si(111)/Au/Ni(8 nm)/Au device [(d), (e)] measured at $V_T$ = -1.4 V, $I_T$ = 1 nA. The STM/BEEM images were recorded at 100 K in (a) and (b) and at RT for (d) and (e). Bright (dark) regions in the BEEM image represents high (low) transmission. Histograms of the distribution in $I_B$ derived from BEEM images in (b) and (e) are shown in (c) and (f), respectively.}
\end{figure} 
\begin{figure}[t]
\includegraphics[scale=0.46]{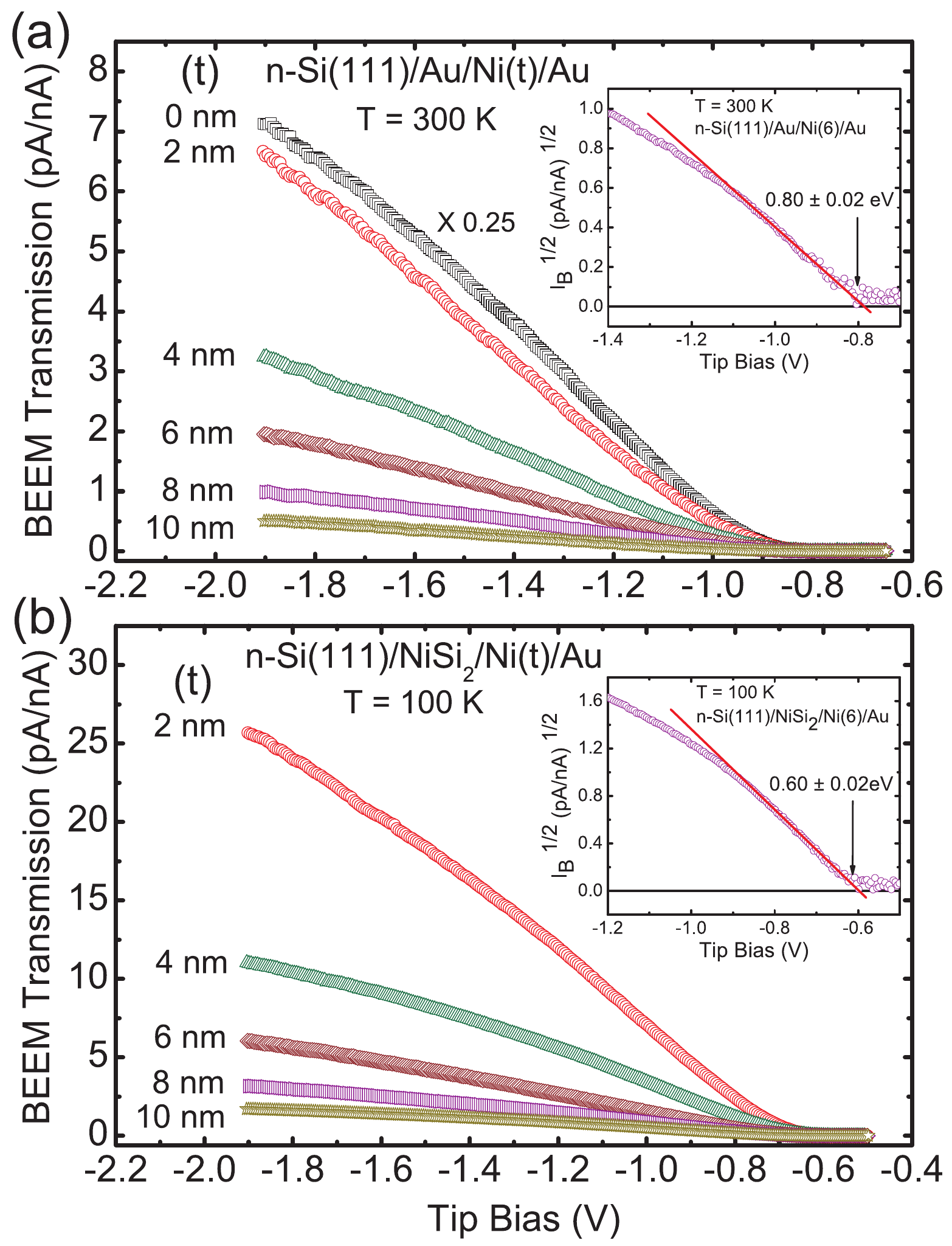}
\caption{\label{fig:BEEM} (a) BEEM transmission per nanoampere of injected tunnel current versus $V_T$ for n-Si(111)/Au(8 nm)/Ni(t)/Au(4 nm), with Ni thicknesses of 0, 2, 4, 6, 8, and 10 nm. The curve for zero Ni thickness has been divided by 4. Measurements were done at RT, and the average SBH found to be 0.80 $\pm$ 0.02 eV, as shown in the inset. (b) BEEM transmission per nanoampere of injected tunnel current versus tip voltage for n-Si(111)/NiSi$_2$(4 nm)/Ni(t)/Au(4 nm) for Ni thicknesses of 2, 4, 6, 8, and 10 nm. Measurements were done at 100 K, and the average SBH found to be 0.60 $\pm$ 0.02 eV, as shown in the inset.}
\end{figure} 
\indent Figure 4 shows the BEEM transmission in Ni, of varying thicknesses, recorded as a function of tip bias $V_T$ and at a constant tunnel current $I_T$. Each spectrum is an average of more than 100 individual spectra taken at several different locations of the device structure. Figure 4(a) represents hot-electron transmission in Ni across the polycrystalline Si(111)/Au interface as well as for a device structure with no Ni layer. Figure 4(b) represents transmission across the epitaxial Si(111)/NiSi$_2$ interface. In both cases, each spectrum was fitted to the BK model \cite{KaiserBell} and the local SBHs extracted by plotting the square root of normalized I$_B$ with V$_T$ as  
\begin{equation}
\sqrt {\frac{I_{B}}{I_T}} {\hspace{0.6 mm}} {\propto} {\hspace{0.6 mm}}(V_T-\phi_B).
\end{equation}
Near the threshold, we find $\phi_B$ at the Si(111)/Au interface to be 0.80$\pm$0.02 eV and 0.60$\pm$0.02 eV at the Si(111)/NiSi$_2$ interface. This matches well with the results from the macroscopic I-V measurement shown in Fig. 2. The anisotropic etching process using 40$\%$ NH$_4$F yields an uniform interface of NiSi$_2$ on Si(111) with an uniquely defined SBH. Figure 4(a) shows that the insertion of a thin (2 nm) Ni layer in Si(111)/Au structure reduces the transmission by a factor of 4. Further, we find that the BEEM transmission is smaller on the polycrystalline Si(111)/Au interface as compared to that on epitaxial Si(111)/NiSi$_2$ for all Ni thicknesses measured.\\
\indent To determine the hot-electron attenuation length ($\lambda$) in Ni, the BEEM transmission at V$_T$ = -1.2 V is plotted with varying Ni thicknesses ($t_{Ni}$) in Fig. 5(a). The solid lines are a fit to the exponential decay of $I_B$ with $t_{Ni}$ as 
\begin{equation}
\frac{I_{B}}{I_T} {\hspace{0.6 mm}}={\hspace{0.6 mm}}R(E)\times exp\left(-\frac{t_{Ni}}{\lambda_{Ni}(E)}\right),
\end{equation}
where R(E) is the energy-dependent proportionality constant and represents the transmittance for zero Ni thickness. The attenuation length in Ni extracted in this way was found to be 3.2 $\pm$ 0.3 nm at -1.2 V for the polycrystalline Si(111)/Au Schottky interface [R(E) = 3.1] and 3.1 $\pm$ 0.3 nm for the epitaxial Schottky interface of Si(111)/NiSi$_2$ [R(E) = 22.0]. By extrapolating $I_B$ to $t_{Ni}$ = 0 [data from Fig. 4 (a)], we find it to be attenuated by a factor of 2.2 due to the addition of two similar Au/Ni interfaces. Interface attenuation arises due to the mismatch of the electron states at both sides of the interface and elastic scattering due to interface disorder, defects, etc. The values of $\lambda_{Ni}$ are then extracted similarly at various STM tip biases for both the device structures and plotted in Fig. 5(b). We find that the attenuation length in Ni does not change significantly with increasing STM tip bias for both device structures. This energy dependence can be correlated to the almost constant density of unoccupied states in Ni into which the hot electrons can propagate. \cite{Bookbandstructure, Zhukov} The attenuation length in Ni determined here is larger than an earlier report measured directly on  Si. \cite{Lu}\\
\begin{figure}[t]
\includegraphics[scale=0.45]{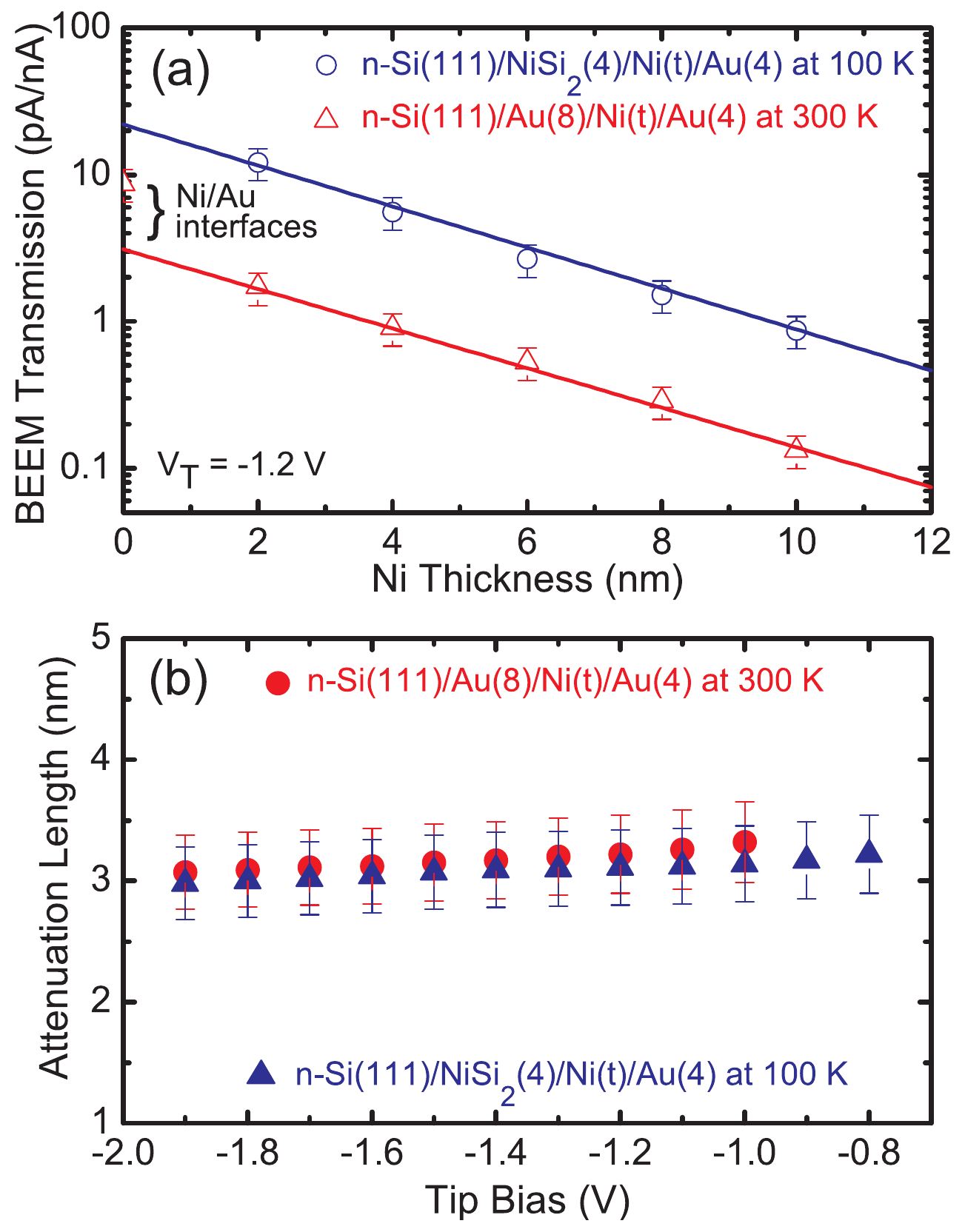}
\caption{\label{fig:BEEM} (a) Normalized BEEM transmission as a function of Ni thickness at $V_T$ = -1.2 V for both the Schottky interfaces. (b) Variation of the attenuation length in Ni with energy for both the Schottky interfaces.}
\end{figure}
\section{\label{sec:level4}IV. Discussion}
\indent To understand the reduction in BEEM transmission at the polycrystalline Schottky interface as compared to that across the epitaxial interface, while $\lambda$ in Ni is the same for both, we consider the different scattering processes relevant for hot-electron transport. Using Matthiessen's rule, the hot-electron attenuation length can be written as the sum of inelastic [$\lambda_i(E)$] and elastic ($\lambda_e$) scattering lengths as  
\begin{equation}
\frac{1}{\lambda(E)} = \frac{1}{\lambda_e} + \frac{1}{\lambda_i(E)}.
\end{equation}
The collected current $I_B$ is a cumulative effect of the hot-electron transmission in Au, Ni and NiSi$_2$ layers, elastic scattering at the Ni/Au and Ni/NiSi$_2$ interfaces, and the transmission probability at the Si/Au and Si/NiSi$_2$ Schottky interface. The hot-electron transmission through a multilayer structure can be described as
\begin{equation}
T \propto {\hspace{0.6 mm}} T_{Au {\hspace{0.6 mm}} or {\hspace{0.6 mm}} NiSi_2}\times T_{Ni}(t)\times T_{Au}.
\end{equation}
The transmissions T$_{Au}$, T$_{Ni}$, T$_{NiSi_2}$ depend exponentially on the hot-electron attenuation lengths ($\lambda$) and the individual film thicknesses as described by Eq. (3). Assuming that the injected hot electrons are similarly attenuated in the top Au and Ni layers for both cases, we now analyze the other contributions to the observed difference in $I_B$. In a recent study, the hot-electron attenuation length $\lambda$ in NiSi$_2$ at V$_T$ = -1.4 V \cite{SP3}, is determined to be 12 nm, similar to that in Au.\cite{SanderAu} We have used a thinner NiSi$_2$ (4 nm) film than Au (8 nm) at the Schottky interface; however, we find from a simple calculation that $I_B$ would still be larger at the epitaxial interface, even for an 8 nm NiSi$_2$ film, using the values of $\lambda$ as stated above. We consider next the attenuation on both sides of the Schottky interface for both Si(111)/Au/Ni and Si(111)/NiSi$_2$/Ni. The good match of the electronic band structure at the Ni/Au interface is expected to result in minimal attenuation of the hot electrons at this interface.\cite{Bookbandstructure} In the absence of band-structure calculation for NiSi$_2$, we cannot analyze the interface attenuation at the Ni/NiSi$_2$ interface. However, the lattice mismatch between Ni and NiSi$_2$ is larger than between Au and Ni, thus qualitatively suggesting that interface attenuation is larger in the former. In spite of the above interface attenuation in the base layers (viz. Au/Ni/Au and Au/Ni/NiSi$_2$), the BEEM transmission for the entire device structure is found to be always larger for the epitaxial interface. This can be understood by the differences in the transmission probability across the Schottky interface. The epitaxial Si(111)/NiSi$_2$ Schottky interface will have a higher transmission probability as compared to the polycrystalline Si(111)/Au Schottky interface, where momentum mismatch of the transmitted hot electrons with that of the available states in Si could lead to a large reduction in the BEEM transmission. \cite{Stiles}\\
\indent The similar values of the hot-electron attenuation length in Ni for both interfaces, while $I_B$ is different, thus points to the fact that $\lambda$ strongly depends on the electron-electron inelastic scattering in bulk Ni and is less sensitive to elastic scattering or other momentum scattering events in the entire device structure. As described earlier in the text, the measurement temperature for both cases is not the same (300 K for the polycrystalline interface and 100 K for the epitaxial interface). To rule out the influence of temperature to the extracted $\lambda$ values, we also performed BEEM transmissions for the polycrystalline interface at 100 K. Although the electron-electron inelastic scattering is temperature independent \cite{Bell}, electron-phonon scattering is not. At 100 K, no significant changes to the BEEM transmission was found, thus suggesting that the influence of acoustic-phonon scattering on the hot electron transmission is insignificant. Our finding of almost similar $\lambda$ in Ni for different Schottky interfaces is different from that in Ref.\cite{Jansen} where a factor of two difference in $\lambda$ was found in Co, for identical Schottky interfaces but different injection interfaces.\\
\section{\label{sec:level5}V. Conclusions}
\indent Our experiments reveal interesting insights into hot-electron transport in ferromagnetic Ni and the relative contribution of the different scattering processes to the extracted hot-electron attenuation length in Ni. By changing the Schottky interface at the collector while keeping the other layers and their thicknesses the same, we see that the overall BEEM transmission is reduced for the polycrystalline interface. The hot-electron attenuation length in Ni, however, does not change significantly across both the interfaces for all the energies measured. This work underpins the dominant contribution of inelastic scattering to the hot-electron attenuation length in bulk Ni for both the device structures. This work will be extended to the study of hot-electron transmission across other epitaxial Schottky interfaces and different combinations of ferromagnetic materials.\\ 
\section{\label{sec:level3}Acknowledgments}
\indent Financial support from the Netherlands Organization for Scientific Research (NWO-VIDI program), the Foundation for Fundamental Research on Matter (FOM) and the NanoNed program coordinated by the Dutch Ministry of Economic Affairs are acknowledged.

\end{document}